\documentclass[aps,prb,superscriptaddress,showpacs,showkeys,floatfix,reprint]{revtex4-1}
\usepackage{graphicx}
\usepackage{dcolumn}
\usepackage[pdftex]{color}
\usepackage{amsmath}
\usepackage{caption}
\usepackage{subcaption}
\usepackage{bm}
\usepackage{gensymb}
\usepackage{verbatim}
\definecolor{darkgreen}{rgb}{0.133,0.545,0.133}
\definecolor{orange}{rgb}{1.0,0.76,0.02}

\begin{document}

%\title{Discovery of new stable and high-temperature Ti-Ta-X shape memory alloys from first principles calculations}
%\title{Discovery of new stable and high-temperature Ti-Ta-X shape memory alloys from first principles calculations by suppression of $\omega$ formation}
\title{Discovery of $\omega$-free high-temperature Ti-Ta-X shape memory alloys from first principles calculations}

\author{Alberto Ferrari}
\email{alberto.ferrari@rub.de}
\affiliation{Interdisciplinary Centre for Advanced Materials Simulation, Ruhr-Universit{\"a}t Bochum, 44801 Bochum, Germany}
\author{Alexander Paulsen}
\affiliation{Institut f{\"u}r Werkstoffe, Ruhr-Universit{\"a}t Bochum, 44801 Bochum, Germany}
\author{Dennis Langenk\"amper}
\affiliation{Institut f{\"u}r Werkstoffe, Ruhr-Universit{\"a}t Bochum, 44801 Bochum, Germany}
\author{David Piorunek}
\affiliation{Institut f{\"u}r Werkstoffe, Ruhr-Universit{\"a}t Bochum, 44801 Bochum, Germany}
\author{Christoph Somsen}
\affiliation{Institut f{\"u}r Werkstoffe, Ruhr-Universit{\"a}t Bochum, 44801 Bochum, Germany}
\author{Jan Frenzel}
\affiliation{Institut f{\"u}r Werkstoffe, Ruhr-Universit{\"a}t Bochum, 44801 Bochum, Germany}
\author{Jutta Rogal}
\email{jutta.rogal@rub.de}
\affiliation{Interdisciplinary Centre for Advanced Materials Simulation, Ruhr-Universit{\"a}t Bochum, 44801 Bochum, Germany}
\author{Gunther Eggeler}
\affiliation{Institut f{\"u}r Werkstoffe, Ruhr-Universit{\"a}t Bochum, 44801 Bochum, Germany}
\author{Ralf Drautz}
\affiliation{Interdisciplinary Centre for Advanced Materials Simulation, Ruhr-Universit{\"a}t Bochum, 44801 Bochum, Germany}

\date{\today}

\begin{abstract}
The rapid degradation of the functional properties of many Ti-based alloys is due to the precipitation of the $\omega$ phase.
In the conventional high-temperature shape memory alloy Ti-Ta the formation of this phase compromises completely the shape memory effect and high ($>100$\degree C) transformation temperatures cannot be mantained during cycling.
A solution to this problem is the addition of other elements to form Ti-Ta-X alloys, which often modifies the transformation temperatures; due to the largely unexplored space of possible compositions, very few elements are known to stabilize the shape memory effect without decreasing the transformation temperatures below 100\degree C. In this study we use transparent descriptors derived from first principles calculations to search for new ternary Ti-Ta-X alloys that combine stability and high temperatures. We suggest four new alloys with these properties, namely Ti-Ta-Sb, Ti-Ta-Bi, Ti-Ta-In, and Ti-Ta-Sc.
Our predictions for the most promising of these alloys, Ti-Ta-Sc, are subsequently fully validated by experimental investigations, the new alloy Ti-Ta-Sc showing no traces of $\omega$ phase after cycling.
Our computational strategy is immediately transferable to other materials and may contribute to suppress $\omega$ phase formation in a large class of alloys.
\end{abstract}

%\pacs{88.88.Aa, 88.88.Aa, 88.88.Aa}

%\keywords{Ti-Ta}

\maketitle

\section{Introduction}
%Smart materials, that can respond to variations of the external conditions with a change of their physical properties or operating behaviour, are becoming more and more popular components because they fit size, efficiency, and durability requirements better than conventional materials \cite{Bhattacharya2005}. 
Among the first discovered smart materials, shape memory alloys (SMAs)\cite{Oelander1932, Chang1951, Funakubo1987, Otsuka1998, Duerig1990, Hornbogen1991, Otsuka1999, VanHumbeeck2001, Kumar2008} are nowadays attractive for actuating applications, efficient energy conversion, and flexible medical instruments and implants. 
SMAs  are ferroelastic materials characterized by a thermal memory, the so-called one-way effect (1WE):
if deformed at low temperature, SMAs are able to recover a predetermined shape by heating.

The 1WE is based on a reversible, solid-to-solid martensitic phase transformation between the high temperture phase (austenite), and the low temperature phase (martensite):
heating an SMA from low temperature induces the nucleation and growth of austenite at the austenite start temperature $A_\text{s}$, and, \textit{vice versa}, cooling an SMA from high temperature induces the nucleation and growth of martensite at the martensite start temperature $M_\text{s}$.

The vast majority of the engineering applications of SMAs use Ni-Ti \cite{Buehler1963, Otsuka2005} as base material, because it combines a durable and reversible 1WE with exceptional physical and mechanical properties.
However, the transformation temperatures $A_\text{s}$ and $M_\text{s}$ of this SMA are lower than 100\degree C \cite{Frenzel2010, Frenzel2015}, which limits the opportunities for designing smart material components in hot environments.

A possible alternative to Ni-Ti as high-temperature shape memory alloys (HTSMAs) \cite{Firstov2004,Ma2010} are Ti-Ta alloys\cite{Bagaryatskii1958,Bywater1972,Fedotov1985,Fedotov1986,Buenconsejo2009,Buenconsejo2009a,Buenconsejo2011,Kim2011,Niendorf2014,Niendorf2015,Niendorf2015a, Chakraborty2015, Chakraborty2016,Kadletz2018,Ferrari2018,Ferrari2019a, Ferrari2019}. %, characterized by an excellent workability and good mechanical properties.
In these alloys, the transformation temperatures $A_\text{s}$ and $M_\text{s}$ increase with decreasing Ta concentration $c_\text{Ta}$, and can be as high as 430\degree C when $c_\text{Ta}$ is reduced to 20 at.\%  \cite{Ferrari2018}. 
The 1WE in Ti-Ta is due to a martensitic transformation between the austenitic phase $\beta$, a solid solution of Ti and Ta with a body-centered cubic lattice and spacegroup $Im\bar{3}m$, and the martensitic phase $\alpha''$, 
%also a solution of Ti and Ta 
with an orthorhombic lattice and spacegroup $Cmcm$.%, as detailed in Refs. \cite{Kadletz2018, Ferrari2018}. 

Unfortunately, at Ta concentrations where the transformation temperature is higher than 100\degree C ($c_\text{Ta}<33$ at.\%) the 1WE in Ti-Ta is not stable and the shape recovery strain decreases rapidly to zero after only a few thermal cycles.
The functional degradation of the 1WE in Ti-Ta, as in other $\beta$-Ti alloys \cite{Kim2018}, is caused by nano-precipitation of the $\omega$ phase \cite{Buenconsejo2009,Buenconsejo2009a,Kim2011,Niendorf2014,Niendorf2015,Maier2017}, a detrimental phase with a hexagonal lattice and spacegroup $P6/mmm$.
The microstructural, thermodynamic, and kinetic aspects of the formation of the $\omega$ phase have recently been discussed in terms of a time-temperature-transformation diagram for Ti-Ta \cite{Paulsen2018}.
The rate of nucleation of the $\omega$ phase is observed to be lower at higher $c_\text{Ta}$, but the formation of this phase in Ti-Ta cannot be avoided unless $c_\text{Ta}$ is increased until $A_\text{s}$ and $M_\text{s}$ become lower than 100\degree C, a regime in which Ni-Ti is usually preferred for engineering applications.

It has been observed in experiments \cite{Buenconsejo2009a, Buenconsejo2011, Kim2011, Zheng2013} that alloying Al, Sn, or Zr  in moderate ($\leq 5$~at.\%) concentrations to Ti-Ta stabilizes the 1WE while, for a specific range of $c_\text{Ta}$, $A_\text{s}$ and $M_\text{s}$ remain higher than 100\degree C.
If there are other alloying elements that can prevent the formation of the $\omega$ phase without decreasing the transformation temperatures to below 100\degree C is an open question of great relevance for alloy design.

In a previous study~\cite{Ferrari2018} we have shown that the transformation temperatures can have a non-intuitive dependence on the Ta and alloying element X concentrations $c_\text{Ta}$ and $c_\text{X}$; 
the same may be true for the free energies of the $\beta$ and $\omega$ phases, that determine the stability of the 1WE.
Therefore, the search for new alloying elements requires experiments that cover the entire $(c_\text{Ta}, c_\text{X})$ space, which are very time consuming if a large number of bulk samples with constant composition have to be manufactured.
This naturally calls for atomistic simulations to guide the design of new Ti-Ta-X alloys.

%The transformation temperatures are determined by the free energies of the austenitic and martensitic phases, whereas the stability of the shape memory effect is given by the free energy difference between the austenitic phase and the detrimental $\omega$ phase.
An accurate estimate of the free energies of the $\alpha''$, $\beta$, and $\omega$ phases in the $(c_\text{Ta}, c_\text{X})$ space with first-principles calculations is perhaps as inefficient as performing the corresponding experiments; %, mainly because of the difficulty to %
%assess all entropy contributions with high precision.
%include exactly the entropy contributions.
%The accuracy in the determination of the free energies must therefore be balanced with computational efficiency.
a trade-off between accuracy and efficiency can be achieved with the derivation of meaningful models that describe to a sufficiently robust approximation the transformation temperatures and the stability, and are based only on information that can be readily extracted from relatively inexpensive first principles data (e.g.~total energies, lattice parameters, densities of states, elastic constants, ...).
%For example, previous computational studies on the binary alloy Ti-Ta have considered only the 0 K energy differences between the $\beta$, $\alpha''$, and $\omega$ phases \cite{Chakraborty2015}, or a simplified treatment of the vibrational entropy within the Debye model \cite{Chakraborty2016} to access the transformation temperatures and stability ranges.

In this article we propose simple and physically motivated descriptors to predict new materials that combine a stable 1WE and transformation temperatures higher than 100\degree C.
By analyzing the relative stability of the $\alpha''$, $\beta$, and $\omega$ phases as a function of $c_\text{Ta}$ and $c_\text{X}$, we shortlist a set of potential candidate materials to a few promising alloys.
%We then test experimentally one of these alloys at the compositions for which the theory predicts a stable and high-temperature 1WE and we validate the results of our model.
We have been able to manufacture one of the alloys, Ti-Ta-Sc, in the composition range for which the theory predicts a stable high-temperature 1WE. % The experimental results validate clearly the predictions of our theoretical model:
In Ti-Ta-Sc the transformation temperatures are higher than 100\degree C and the $\omega$ phase is completely absent from the sample after thermal cycling, resulting in a remarkable improvement of the stability of the shape memory effect with respect to Ti-Ta, in full agreement with the predictions.  

The precipitation of the $\omega$ phase presents a long-standing technological challenge in Ti-base alloys in general\cite{Hickman1969}. Over the years and based on experience and insight, alloy constituents such as Al, Zr, O, etc.~or microstructral features were associated to the suppression of $\omega$ phase formation \cite{Qazi2005,Talling2009,Guo2013,Pang2018,Tane2019}. Our computational design strategy may be readily applied to evaluate the relative stability of the $\omega$ phase in Ti-alloys, without ad-hoc or experience-based assumptions on particular alloying elements, and may give important impetus towards the rational design of Ti-based alloys.

\section{Methods}

\subsection{Computational Setup}

The first principles calculations have been performed using density functional theory with the plane-waves pseudopotential code VASP 5.4 \cite{Kresse1993,Kresse1996,Kresse1996a}.
The recommended projector augmented wave (PAW) pseudopotentials \cite{Bloechl1994,Kresse1999} with the PBE expression \cite{Perdew1996} for the exchange correlation functional have been employed for all the elements.
The energy cutoff has been fixed to 450 eV and the k-point meshes, distributed according to the Monkhorst-Pack scheme \cite{Baldereschi1973,Monkhorst1976}, have been set to $(6\times8\times8)$ and  $(5\times5\times9)$ for the $3\times2\times2$ and $2\times2\times2$ supercells of the $\beta$ and $\omega$ phases, respectively, and to $(8\times6\times6)$ for the orthorhombic supercells considered in the binary interaction method\cite{Ferrari2018} (see Sec.~\ref{Bin_Int_M}). 
The metallic electronic occupations have been smeared with the Methfessel-Paxton function \cite{Methfessel1989} of order 1 with a width of 0.05 eV.
Since Ti-Ta-X alloys are solid solutions, we have evaluated the relative stability of the $\beta$ and $\omega$ phases using special quasirandom structures (SQS)\cite{Zunger1990}, small supercells that best represent the spatial n-body correlations of random structures.
The SQS were generated with a Metropolis Monte Carlo program derived from the ATAT package \cite{Pezold2010,Kossmann2015,VanDeWalle2002} taking into account the spatial correlations up to five body terms.
The energy at the equilibrium volume has been computed with a Birch-Murnaghan equation of state fit\cite{Murnaghan1944,Birch1947} after complete relaxation of the atomic degrees of freedom and the cell shape.
All calculations presented in this work were spin-unpolarized, as test calculations including spin-polarization for structures involving Co and Ni converged to non-magnetic states. 

\subsection{Stability of the 1WE}
Since the $\omega$ phase is observed to form from austenite, we have computed the 0 K energy difference between the $\beta$ and $\omega$ phases $\Delta E^{(\beta-\omega)} (c_\text{Ta},c_\text{X})$ to describe the stability of the 1WE.
If the compositional dependence of the entropy for the phase transition $\beta \rightarrow \omega$ is neglected, then $\Delta E^{(\beta-\omega)} (c_\text{Ta},c_\text{X}) \leq 0$ signals a region in the $(c_\text{Ta},c_\text{X})$ space where the formation of the detrimental $\omega$ phase is unfavorable.

For each potential SMA, we have fixed the composition of the alloying element X to 4~at.\%, and calculated the Ta concentration at which $\Delta E^{(\beta-\omega)} (c_\text{Ta},c_\text{X}=4 \, \text{at}.\%)=0$. 
Since this concentration is also known for pure Ti-Ta \cite{Chakraborty2015}, we then linearly interpolated the \textit{locus} for which $\Delta E^{(\beta-\omega)} (c_\text{Ta},c_\text{X}) = 0$.
This line separates the region in which $\omega$ is more stable than $\beta$ from the region in which $\beta$ is more stable than $\omega$.

%for three different values of $c_\text{Ta}$ and then linearly extrapolated the Ta concentration $c^\star_\text{Ta}$ at which
%\begin{equation}
%\Delta E^{(\beta-\omega)} (c^\star_\text{Ta},c_\text{X}=4 at.\%) = 0, 
%\end{equation}
%that is, $c^\star_\text{Ta}$ is such that for $c_\text{Ta}\geq c^\star_\text{Ta}$ the alloy Ti-Ta-4X is stable.
%The linear dependence of $\Delta E^{(\beta-\omega)} (c_\text{Ta},c_\text{X})$ as a function of $c_\text{Ta}$ is justified by the exceptionally low mixing energy of Ti and Ta \cite{Ferrari2018}.

\subsection{Transformation temperatures}
\label{Bin_Int_M}
To determine the region in the $(c_\text{Ta},c_\text{X})$ space where the transformation temperatures are higher than 100\degree C, we have calculated the 0 K energy difference between the $\beta$ and $\alpha''$ phases $\Delta E^{(\beta-\alpha'')}(c_\text{Ta},c_\text{X}) $.
In Ti-Ta based alloys, the entropy difference between the two phases depends very weakly on $c_\text{Ta}$ and $c_\text{X}$ \cite{Ferrari2019} and can be assumed to be constant.
Therefore, the 0 K energy difference is usually sufficient to estimate the much more computationally expensive free energy difference between austenite and martensite.

In our previous work \cite{Ferrari2018} we have shown that to a first approximation Ti-Ta-X alloys can be treated as ideal solid solutions for which $\Delta E^{(\beta-\alpha'')}$ takes the form
\begin{equation}
\Delta E^{\beta-\alpha''} \simeq \text{A} \cdot c_\text{Ta} + \text{B} \cdot c_\text{X}+ \text{C} \cdot c_\text{Ta} c_\text{X} + \text{D},
\label{fitting_function}
\end{equation}
where A and D are parameters that depend on Ti and Ta, and B and C are parameters that depend on the interaction of the alloying element X with Ti and Ta.

To quickly estimate  the coefficients B and C we have employed the binary interaction method that we have presented in Ref.~\onlinecite{Ferrari2018}.
According to this approach, B and C can be calculated from the energy difference between the $\beta$ and $\alpha''$ phases of pure Ti, pure X, and artificial binary Ti-X and Ta-X solid solutions.

In fact, the mixing energy of a system with elements of species n ($\text{n}=\text{Ti, Ta, X}$) in the phase (i), defined as
\begin{equation}
 ^\text{mix}E^\text{(i)}:= E^\text{(i)} - \sum_\text{n} c_\text{n} E^\text{(i)}_\text{n},
\label{mixing}
\end{equation}

can be expanded as\cite{Ferrari2018}

\begin{equation}
 ^\text{mix}E^\text{(i)} =  \sum_\text{n,m} k^\text{(i)}_\text{nm} c_\text{n} c_\text{m} +O[c^3]
 \label{mixingexp}
\end{equation}

for regular solid solutions.
As detailed in Ref.~\onlinecite{Ferrari2018}, the quadratic coefficients $k^\text{(i)}_\text{nm}$ of this expansion for the $\beta$ and $\alpha''$ phases can be fitted from the mixing energies of binary n-m solid solutions. 
If the difference between these quadratic coefficients is expressed as

\begin{equation}
\Delta k_\text{nm}:= k^{(\beta)}_\text{nm}-k^{(\alpha'')}_\text{nm}
\end{equation}

and furthermore

\begin{equation}
\lambda_\text{n}:=E^{(\beta)}_\text{n}-E^{(\alpha'')}_\text{n} \quad ,
\end{equation}

then the coefficients B and C are simply given by\cite{Ferrari2018}:

\begin{align}
\label{system}
\text{B}&=\Delta k_\text{TiX}+\lambda_\text{X}-\lambda_\text{Ti}\\ \nonumber
\text{C}&=\Delta k_\text{TaX}-\Delta k_\text{TiX} \quad .\\ \nonumber
\end{align}

With the binary interaction method, it is possible to calculate the energy difference $\Delta E^{\beta-\alpha''}$ in the entire composition range using eq.~\eqref{fitting_function}. 
This  considerably reduces the computational cost associated with the estimation of the transformation temperatures in the $(c_\text{Ta}, c_\text{X})$ space.

The approximations underlying this approach derive from the truncation of the expansion in eq.~\eqref{mixingexp} and from the fitting of the coefficients $k^\text{(i)}_\text{nm}$, that can be biased by the fact that not every pair of elements n-m can form solid solutions in a specific phase (i).

\subsection{Experimental Setup}

The Ti-Ta-Sc sample has been prepared by arc melting high purity Ti, Ta, and Sc raw materials.
The SMA ingot has been remelted 15 times to achieve chemical homogeneity. 
The actual composition of the ingot has been measured by energy dispersive X-ray analysis (EDX) in a scanning electron microscope and determined as approximately $\text{Ti}_{66.2}\text{Ta}_{31.5}\text{Sc}_{2.3}$. 
Details on the thermomechanical processing and the chemical analysis are given in Refs.~\onlinecite{Zhang2014,Frenzel2015}.

The fully recrystallized alloy has been subjected to thermal cycling in a differential scanning calorimetry instrument of type TA 2920 CE.
Details on DSC operating parameters are given in Refs.~\onlinecite{Frenzel2015,Paulsen2018}.

To obtain electron-transparent samples for the TEM microstructure analysis, a focused ion beam system of type FEI Helios G4 CX DualBeam has been used.
The TEM characterization has been conducted on a Tecnai F20 G2 Supertwin FEG TEM, operating at an acceleration voltage of 200 kV. 
All further details on TEM sample preparation and analysis are available in Refs.~\onlinecite{Zhang2014,Niendorf2015,Paulsen2018,Langenkamper2019}.

\section{Results and discussion}

\begin{figure}
\begin{centering} 
\includegraphics[scale=0.15]{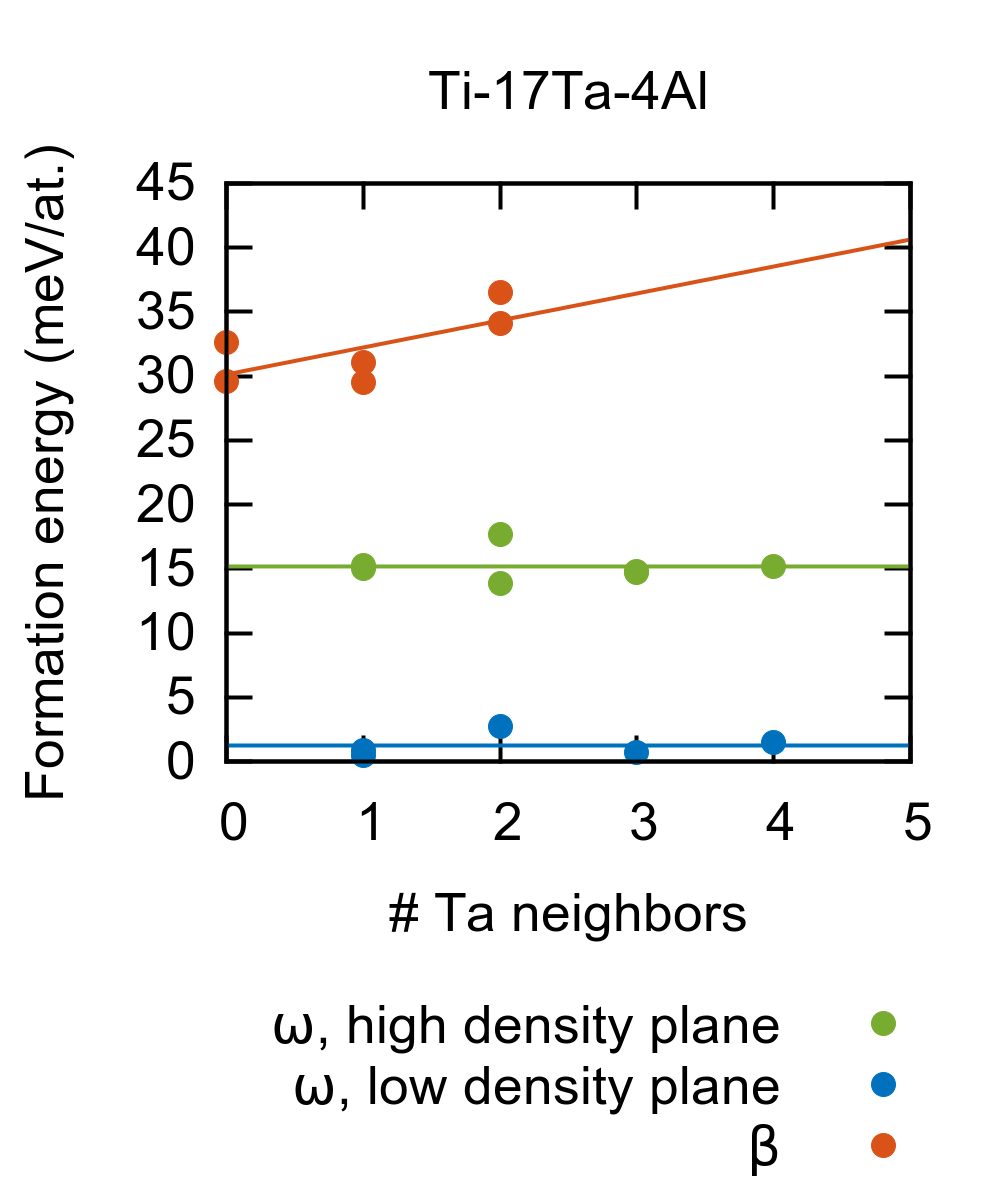}
\par\end{centering}
\caption{Formation energy of Ti-17Ta-4Al in the $\beta$ and $\omega$ phases. For the $\omega$ phase, the formation energy is different if Al occupies the Wyckoff position in the high or in the low density plane.}
\label{Ti-17Ta-4Al}
\end{figure}

\begin{figure*}
\begin{centering} 
\includegraphics[scale=0.18]{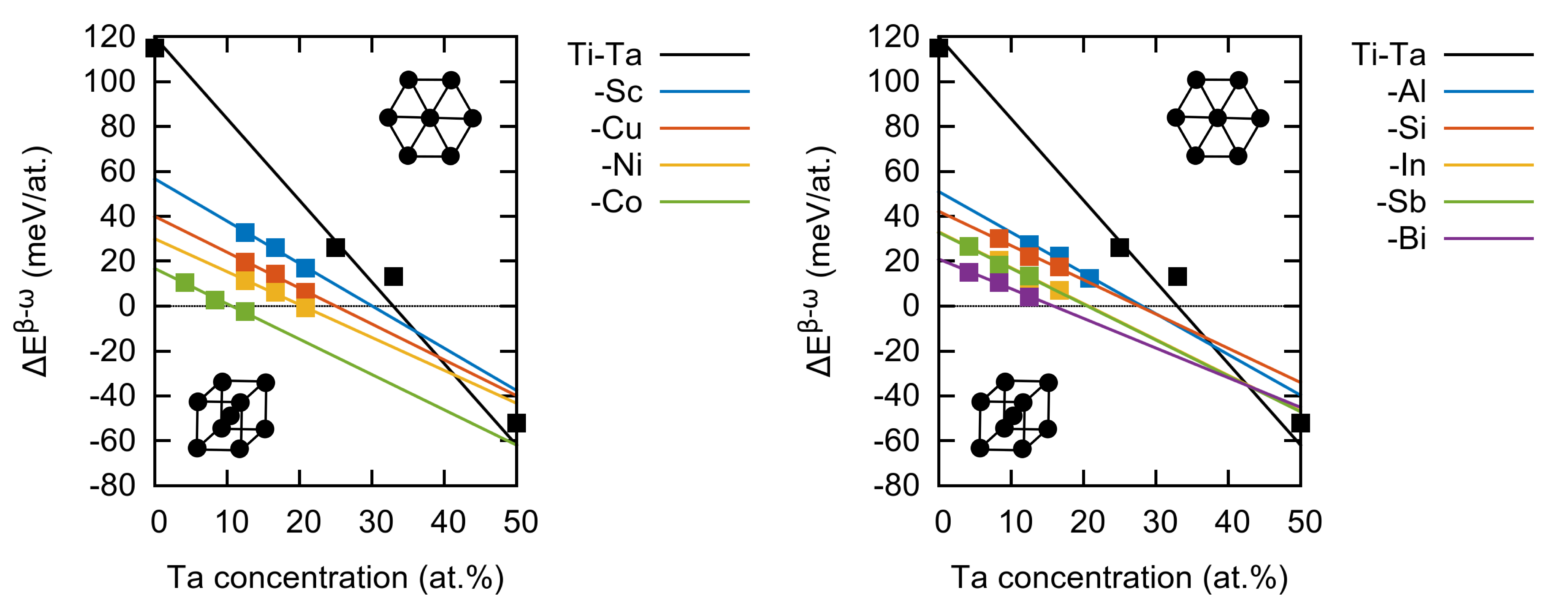}
\par\end{centering}
\caption{Energy difference between $\beta$ and $\omega$ as a function of  Ta concentration for Ti-Ta-4X alloys, where X is a $d$-valent (left) or $p$-valent (right) element.}
\label{omega}
\end{figure*}

We have restricted our search for possible alloying elements to transition metals and $p$-valent metals.
From these candidates we have excluded noble gases, noble metals, poisonous or radioactive elements, and the 2$p$ elements, which are more likely to occupy interstitial sites rather than producing substitutional defects.
Some of the remaining elements, namely Al, V, Cr, Fe, Zr, Mo, Sn, and Hf, have been already investigated by Buenconsejo \textit{et al.} \cite{Buenconsejo2009a}, albeit at a fixed Ta concentration.
We have chosen to study eight new elements, Si, Sc, Co, Ni, Cu, In, Sb, and Bi, as possible candidates for alloying elements in new SMAs.
To benchmark our approach, we have chosen to analyze also Ti-Ta-Al, for which detailed experimental results are already present in the literature \cite{Buenconsejo2011, Ferrari2018}.

\bigskip

For each alloying element, we have investigated the site preference of the substitutional defects in both the $\beta$ and $\omega$ phases.
An example of the formation energies of these two phases, defined as
\begin{equation}
E^\text{(i)}_f:= E^\text{(i)} - c_\text{Ti} E_\text{Ti} - c_\text{Ta} E_\text{Ta} - c_\text{X} E_\text{X} \quad ,
\label{form}
\end{equation}

where $E^\text{(i)}$ is the total energy of Ti-Ta-X in  phase (i) ($\beta$ or $\alpha''$), $E_\text{Ti}$ is the energy of hcp Ti, $E_\text{Ta}$ of bcc Ta, and $E_\text{X}$ of the most stable structure of the element X, is displayed in Fig.~\ref{Ti-17Ta-4Al} for Ti-17Ta-4Al (with 17 at.\% Ta and 4 at.\% Al).
The $x$-axis corresponds to the number of Ta atoms in the first nearest neighbor shell of Al.
For $\beta$, the formation energy increases with an increasing number of Ta nearest neighbors for all the investigated alloys, as already noted for Ti-Ta-Al, Ti-Ta-Sn, and Ti-Ta-Zr \cite{Ferrari2018}, apart from Ti-Ta-Sc, for which it is approximately constant.
For $\omega$, the formation energy is instead independent of the number of Ta nearest neighbors, but depends on the Wyckoff site in which the alloying element is positioned:
$\omega$ is characterized by three sites, two of which are equivalent to each other and located on a high density plane perpendicular to the [0001] direction, and the other one on a low density plane perpendicular to the same direction.
The elements with an atomic radius larger than Ti (Al, Sc, In, Sb, and Bi) show a site preference for the low density plane, whereas the elements with an atomic radius smaller than Ti (Si, Co, Ni, and Cu) for the high density plane.

As diffusion to the most stable site is kinetically hindered~\cite{Ferrari2018}, we have occupied the lattice sites stochastically.
We have thus assumed $8 \times c_\text{Ta}$ as the number of Ta nearest neighbors in $\beta$ (8 is the number of first nearest neighbors in a bcc structure) and computed the corresponding formation energy.
For the formation energy of $\omega$ we have averaged the formation energies of configurations with substitutions in the three Wyckoff positions.

Fig.~\ref{omega} shows the resulting energy difference $\Delta E^{(\beta-\omega)}$ as a function of $c_\text{Ta}$ with fixed $c_\text{X}= 4$~at.\% for the $d$ (left) and $p$ (right) valent alloying elements.
For comparison, the data for pure Ti-Ta taken from Chakraborty \textit{et al.} \cite{Chakraborty2015} are also reported (black dots).
A negative value of $\Delta E^{(\beta-\omega)}$ indicates a stable 1WE, and the intercept with the x-axis indicates the concentration at which the energies of $\beta$ and $\omega$ are equal for $c_\text{X}=4$ at.\%.

It can be seen that all selected alloying elements destabilize the detrimental $\omega$ phase with respect to $\beta$.
Among the $p$-valent elements there is a clear trend with the size of the alloying element:
elements with higher atomic radii tend to destabilize the $\omega$ phase more.
This can be understood from the fact that configurations with relatively large elements in the high density plane of the $\omega$ phase are energetically very unfavorable.
No clear trend in terms of size or band filling is instead recognized for the $d$-valent elements.

To evaluate the compositional dependence of the transformation temperatures as described by eq.~\eqref{fitting_function}, we have taken the values of A= -23.9 K/at.\% and D=1140~K from our previous work~\cite{Ferrari2018}.
To estimate the coefficients B and C we have fitted the mixing energy of binary Ti-X and Ta-X alloys using eq.~\eqref{mixing} as described in Ref.~\onlinecite{Ferrari2018}.

\subsection{Benchmark: Ti-Ta-Al}

\begin{figure}
\begin{centering} 
\includegraphics[scale=0.15]{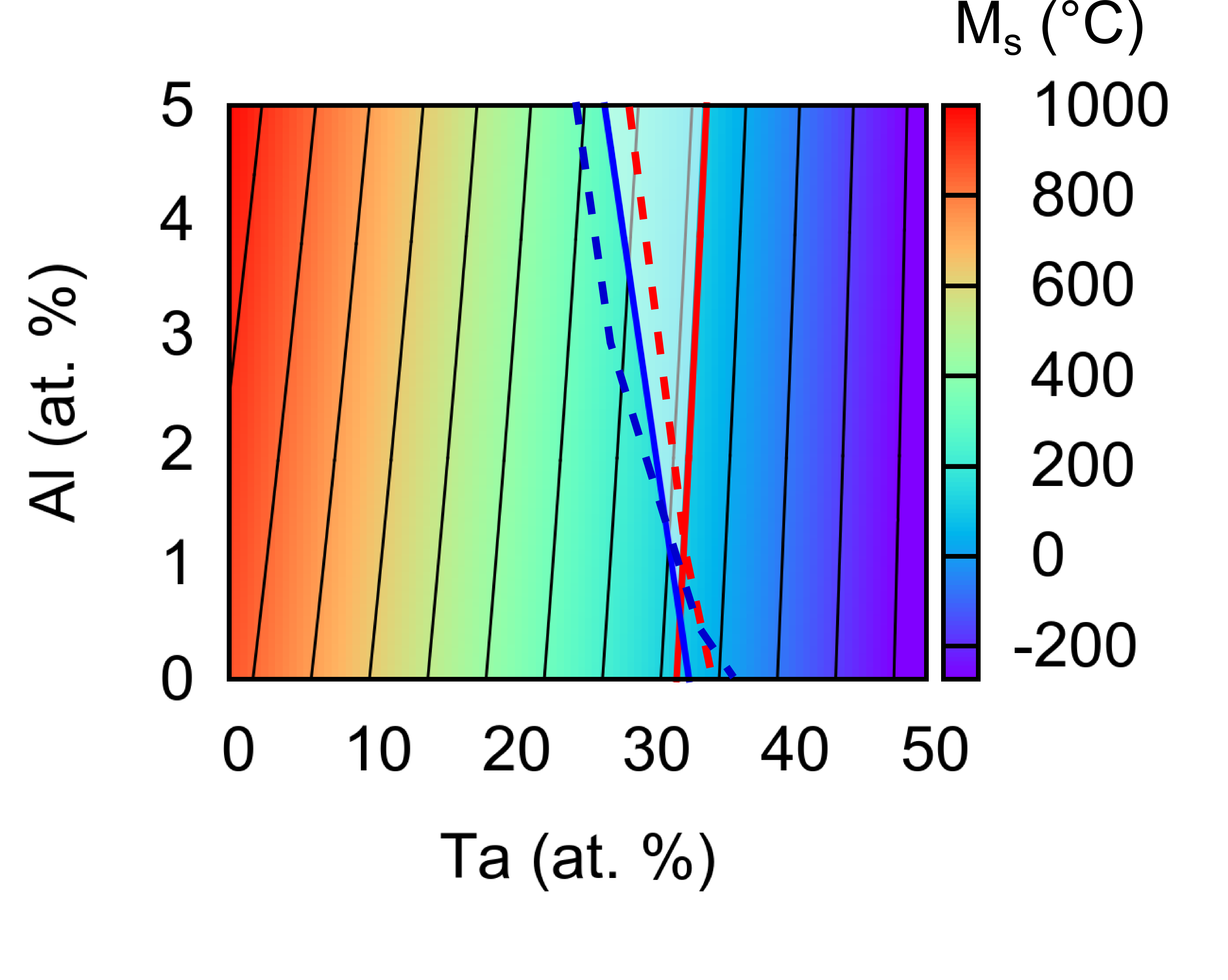}
\par\end{centering}
\caption{Martensitic start temperature predicted by the theoretical model as a function of the Ta and Al concentration in Ti-Ta-Al. The red line divides the region in which $M_\text{s}>100$\degree C (left) from the region in which $M_\text{s}<100$\degree C (right). The blue line separates the region in which the $\omega$ phase is stable (left) from the region in which the $\omega$ phase is unstable (right). The shaded area between the blue and red lines is a region where the 1WE is stable and has $M_\text{s}>100$\degree C. Solid lines are predictions from our model and dashed lines are experimental data from Ref.~\onlinecite{Buenconsejo2011}.}
\label{Ti-Ta-Al}
\end{figure}

The predictions of our model for the stability and high temperature regions in Ti-Ta-Al are reported in Fig.~\ref{Ti-Ta-Al}.
The color scale indicates the predicted $M_\text{s}$ as a function of $c_\text{Ta}$ and $c_\text{X}$ from eq.~\eqref{fitting_function}.
The solid red line separates the predicted regions of high (left) and low (right) $M_\text{s}$, and the blue line separates the predicted regions where $\omega$ is stable (left) and unstable (right).
A region of the $(c_\text{Ta}, c_\text{Al})$ plane delimited with a blue line on the left and a red line on the right is predicted to be characterized by $M_\text{s}>100$\degree C \textit{and} a stable 1WE.
From Fig.~\ref{Ti-Ta-Al} it can be deduced that such a region cannot be obtained in binary Ti-Ta, but only with the addition of Al, in agreement with previous investigations \cite{Buenconsejo2009a,Buenconsejo2011,Niendorf2015}.

The experimental curves \cite{Buenconsejo2011} for the stability (blue) and high-temperature (red) regions are displayed as dashed lines in Fig.~\ref{Ti-Ta-Al} for comparison.
The stability line from our model agrees well with the experimental measurements and the stability/instability regions can be predicted within roughly 3~at.\% Ta. 
The red line from our model has a positive slope, indicating that for increasing Al content $M_\text{s}$ would increase slightly at $c_\text{Ta}\sim30$~at.\%.
As already pointed out in our previous work \cite{Ferrari2018}, this is not in quantitative agreement with experiment, as in Ti-Ta-Al the transformation temperatures have been observed to increase for increasing $c_\text{Al}$ only for $c_\text{Ta}<16$~at.\% \cite{Ferrari2018}.
This is due to the approximations within  the binary interaction model.
Despite this, our model is able to predict qualitatively the existence of a region with high-temperature and stable 1WE and 
%is thus suitable to be applied to the new alloys.
is thus suitable to guide the assessment of new alloys.

\subsection{New candidate alloys}

\begin{figure*}
\begin{centering} 
\includegraphics[scale=0.14]{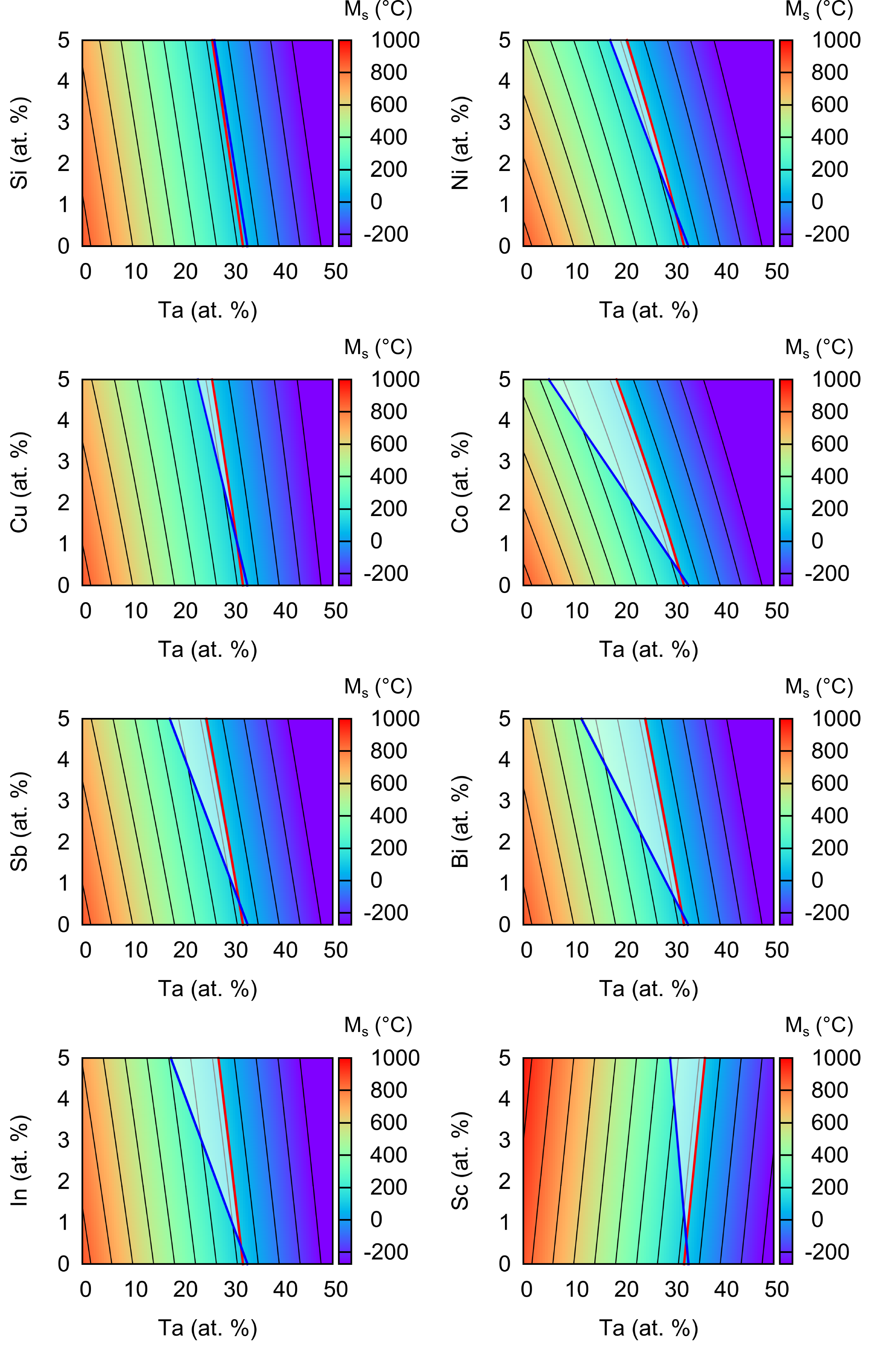}
\par\end{centering}
\caption{Predicted martensitic start temperature as a function of composition for (left to right, top to bottom): Ti-Ta-Si, Ti-Ta-Ni, Ti-Ta-Cu, Ti-Ta-Co, Ti-Ta-Sb, Ti-Ta-Bi, Ti-Ta-In, and Ti-Ta-Sc. The red and blue lines mark the region of high transformation temperature and high stability, respectively, as detailed in Fig.~\ref{Ti-Ta-Al}}%The red frame indicates the most promising alloys according to our predictions.}
\label{maps}
\end{figure*}

Fig.~\ref{maps} shows the predicted diagrams for the stability and martensitic start temperature for Ti-Ta-Si, Ti-Ta-Ni, Ti-Ta-Cu, Ti-Ta-Co, Ti-Ta-Sb, Ti-Ta-Bi, Ti-Ta-In, and Ti-Ta-Sc.
In general, the elements that destabilize the $\omega$ phase the most, like Co and Bi, are found to lower  $M_\text{s}$ considerably, because the $\beta$ phase is strongly stabilized.
For most of the elements a balance between stability and high transformation temperatures can be found by an  appropriate tuning of  $c_\text{Ta}$ and $c_\text{X}$, although some alloys appear to be more promising than others.

In particular, for Ti-Ta-Si we predict no region of stability and high transformation temperature, thus this alloy is unlikely to be a good SMA.
The additions of Ni and Cu result in  very narrow regions of stability and high transformation temperature, with a width comparable to the error bars of our model.
%Nothing can thus be concluded about the performance of Ti-Ta-Ni and Ti-Ta-Cu SMAs.
A definitive conclusion concerning the performance of Ti-Ta-Ni and Ti-Ta-Cu as SMAs is thus not possible.
Alloying Co is predicted to decrease  $M_\text{s}$ substantially, although a region of stability and high transformation temperature can be identified at relatively low $c_\text{Ta}$.
However, the stability at such a low $c_\text{Ta}$ may still be compromised by the precipitation of isothermal $\omega$ particles with a Ti-rich composition \cite{Ferrari2019a};
therefore, we presume that Co may not improve the stability of the 1WE.

The $p$-valent elements Sb, Bi, and In, characterized by a similar chemistry upon alloying to Ti-Ta, might be potential candidates to stabilize the 1WE, although they are predicted to decrease $M_\text{s}$ at all $c_\text{Ta}$.
In particular In, in the same period as Sn and isoelectronic to Al, shares the same beneficial properties of these two elements, already known to favor the stability of the 1WE \cite{Buenconsejo2009a}.

Finally, the alloy Ti-Ta-Sc seems to be superior to the other investigated potential SMAs because the addition of Sc can destabilize the $\omega$ phase while keeping $M_\text{s}$ high even at $c_\text{Ta}\sim30$~at.\%.
Our calculations predict that $M_\text{s}$ should increase slightly with increasing $c_\text{Sc}$.
An increase of the energy difference between $\beta$ and $\alpha''$ has been observed recently also in Ti-Nb-Sc \cite{Minami2017} and can be imputed to a band-filling effect: alloying Sc decreases the number of $d$ valence electrons of the alloy and destabilizes the $\beta$ phase.
%The surprising feature about Ti-Ta-Sc is that
This decrease of the $d$-electron count, however, does not result in a stabilization of the $\omega$ phase but in a destabilization, presumably because of the size mismatch between Ti and Sc:
alloying Sc, with an atomic radius bigger than that of Ti, is not favorable in the $\omega$ phase.
Given the very promising results for this alloy, we decided to investigate  the Ti-Ta-Sc system experimentally.

\subsection{Experimental validation for Ti-Ta-Sc}

\begin{figure*}
\begin{centering}
\includegraphics[scale=0.085]{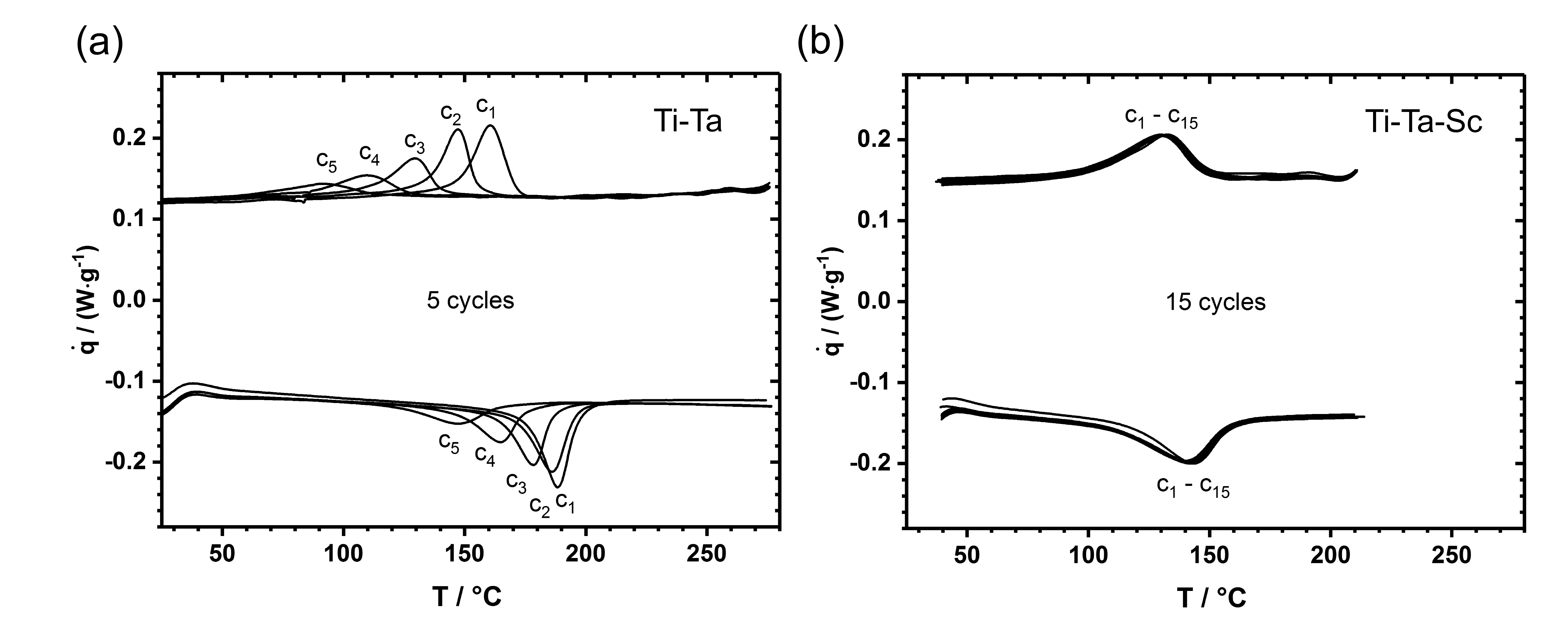}
\par\end{centering}
\caption{
(a) DSC data for 5 heating/cooling cycles of binary Ti-Ta with $c_\text{Ta}=30\text{ at.}\%$, where a rapid degradation can be observed. 
(b) 15 DSC cycles of Ti-Ta-Sc with $c_\text{Ta}=31.5\text{ at.}\%$ and $c_\text{Sc}=2.3\text{ at.}\%$ , which exhibits a remarkably stable behavior and transformation temperatures higher than 100\degree C.
A comparison of the data in (a) and (b) provides a clear evidence for the improved functional stability of Ti-Ta-Sc over Ti-Ta.
}
\label{DSC}
\end{figure*}

\begin{figure}
\begin{centering}
\includegraphics[scale=0.15]{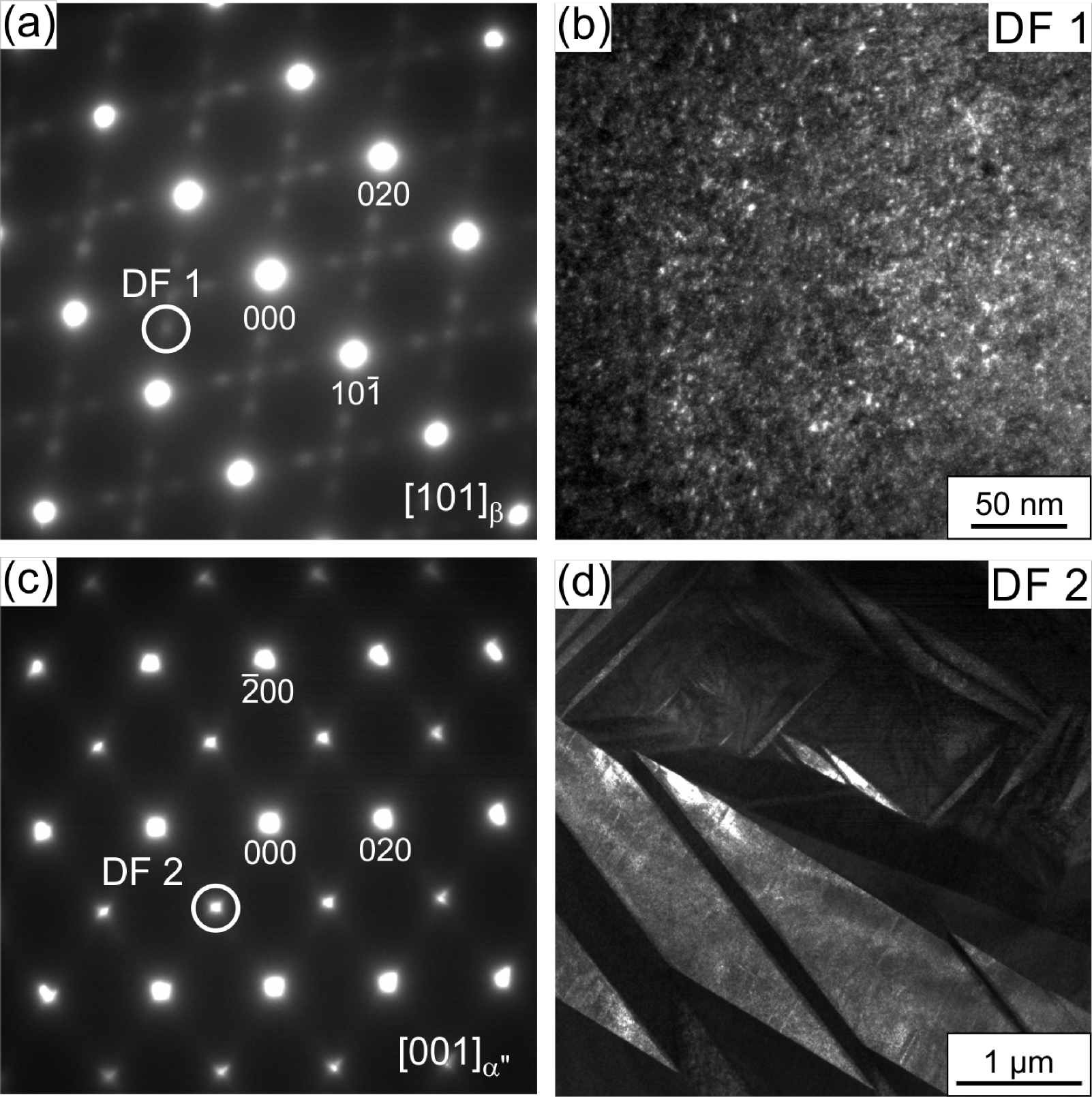}
\par\end{centering}
\caption{TEM analysis of the Ti-Ta and Ti-Ta-Sc samples after thermal cycling.
(a) Selected-area diffraction pattern of the binary Ti-Ta sample at the $[101]_\beta$ zone axis, revealing diffraction intensities of the $\omega$ phase. 
(b) Dark field (DF) image taken for the reflection marked DF1 in (a), indicating the presence of nano-scaled $\omega$ precipitates.
(c) Selected-area diffraction pattern at the $[001]_{\alpha''}$ zone axis of the Ti-Ta-Sc sample, showing no $\omega$ reflections. 
(d) Only martensitic features are identified in the DF image obtained for the $\alpha''$ diffraction intensity marked DF2 in (c).}
\label{TEM}
\end{figure}

To validate the theoretical predictions, we have fabricated a Ti-Ta-Sc alloy with $c_\text{Ta}=31.5\text{ at.}\%$ and $c_\text{Sc}=2.3\text{ at.}\%$, a composition inside the proposed high-temperature and stability region for this alloy, and evaluated the functional and microstructural stability of this alloy with differential scanning calorimetry (DSC) and transmission electron microscopy (TEM).

Figure \ref{DSC} compares the DSC data obtained from thermal cycling experiments on Ti-Ta and Ti-Ta-Sc.
The exothermic peaks on cooling (positive heat flow) indicate the formation of martensite, whereas the endothermic peaks on heating (negative heat flow) are associated with the reverse transformation $\alpha'' \rightarrow \beta$.
Both alloys have been subjected to cyclic heating and cooling and the cycle numbers are marked with c$_i$.

As indicated by the strong shift of the transformation peaks to lower temperatures in Fig.~\ref{DSC}(a), the phase transformation in binary Ti-Ta alloy is not stable and degrades within only 5 cycles.
For the unstable binary alloy, thermal cycling is also associated with a change in the latent heat, which corresponds to the area below the DSC peaks.
The small transformation peaks obtained in the 5th cycle of Ti-Ta suggest that a significantly smaller volume fraction of the material undergoes a martensitic and reverse transformation during cycling.

In contrast, the situation is drastically different in Fig.~\ref{DSC}(b) for the new Ti-Ta-Sc alloy. 
Ti-Ta-Sc shows a very stable transformation behavior, as all heating/cooling curves coincide almost perfectly up to at least 15 cycles.
Furthermore, the martensitic and reverse transformation is observed at a temperature higher than 100\degree C, in agreement with the theoretical predictions.

The microstructures of Ti-Ta and Ti-Ta-Sc after thermal cycling have also been characterized by TEM to identify potential degradation mechanisms.
For binary Ti-Ta, a selected area diffraction pattern at the $[101]_\beta$ zone axis, shown in Fig.~\ref{TEM}(a), reveals strong diffraction intensities at $1/3 \, \langle 211 \rangle_\beta$ positions, associated with the presence of the $\omega$ phase. 
Based on the dark field image in Fig.~\ref{TEM}(b), obtained for the marked $\omega$ reflection, nano-scaled $\omega$ precipitates with a high volume fraction have been identified in Ti-Ta.
%This phase is known to lead to the suppression of the martensitic transformation and therefore to functional degradation on cycling. In addition, $\alpha''$ intensities appear in the diffraction pattern at $1/2 \langle 211 \rangle_\beta$ positions revealing that martensite is not fully suppressed.

Conversely, the selected area diffraction pattern of the Ti-Ta-Sc sample at the $[001]_{\alpha''}$ zone axis in Fig.~\ref{TEM}(c) indicates a purely martensitic matrix and no diffraction intensity corresponding to the $\omega$ phase at $1/3 \, \langle 211 \rangle_\beta$ positions is observed.
Therefore, in agreement with the theoretical predictions, no traces of $\omega$ phase precipitation are detected.
Accordingly, the dark field image in Fig.~\ref{TEM}(d) obtained for the encircled $[\bar{1}10]$ reflection shows a microstructure with typical martensitic features. 
This proves that the addition of Sc to Ti-Ta results in a complete suppression of the detrimental $\omega$ phase.

\bigskip

\bigskip

\section{Conclusions}

%In summary, we have 
We presented a theory-guided alloy optimization of Ti-Ta-X SMAs  that can discover alloy compositions demonstrating a superior stability with respect to thermal cycling and high transformation temperatures.
Our first principles screening, based on
%the calculation of 
0~K energy differences between random structures, has identified at least four potential stable and high temperature SMAs, namely Ti-Ta-Sb, Ti-Ta-Bi, Ti-Ta-In, and Ti-Ta-Sc.
We have experimentally fabricated the most promising of these new alloys, Ti-Ta-Sc, and observed an extremely good stability of the 1WE because of the full suppression of the $\omega$ phase, and transformation temperatures higher than 100 \degree C, in agreement with the predictions of the model.
The ternary alloys described in this study may open new opportunities for the application of SMAs in high temperature environments;
these opportunities are even broadened by the possibility to apply the workflow described here to explore other ternary Ti-Ta-X alloys. 
Our approach is fully transferable to other, even quaternary or multicomponent Ti-alloys and forms the basis for a rational design of $\omega$-free alloys.

\section*{Acknowledgements}
Financial support from the Deutsche Forschungsgemeinschaft (DFG) within the research unit FOR 1766 (High Temperature Shape Memory Alloys, http://www.for1766.de, project number 200999873, sub-groups TP1, TP2, and TP3) is thankfully acknowledged.
Part of the calculations has been performed on the supercomputers of the Swedish National Infrastructure for Computing (SNIC) at the National Supercomputer Centre (NSC) in Link\"oping and of the Center for High Performance Computing (PDC) in Stockholm.

%\section*{Author Contributions}
%A.F.~carried out the DFT calculations and performed the analysis of the numerical data with guidance from J.R.~and R.D. 
%The Ti-Ta and Ti-Ta-Sc samples were fabricated by A.P.~and D.P, and A.P.~performed the DSC measurements. D.L.~and %C.S.~performed the TEM measurements. 
%J.F.~supervised the experiments. 
%J.F.~and G.E.~designed the experimental research plan. 
%J.R. and R.D.~designed the computational research plan and supervised the calculations.
%A.F.~wrote the manuscript with input from J.F., J.R., and R.D.
%All authors discussed and interpreted the results.

%\section*{Data Availability}
%The datasets generated during the current study are available from the corresponding author on request.
%
%\section*{Competing Interests}
%The authors declare no competing financial or non-financial interests.

\bibliography{./bib/biblio}

\end{document}